\documentclass[10pt,journal]{IEEEtran}

\usepackage{cite}
\usepackage{amsmath,amssymb,amsfonts}
\usepackage{algorithmic}
\usepackage{graphicx}
\usepackage{textcomp}
\usepackage{booktabs}
\usepackage{url}

\makeatletter
\g@addto@macro\UrlBreaks{\do\_}
\makeatother
\def\BibTeX{{\rm B\kern-.05em{\sc i\kern-.025em b}\kern-.08em
    T\kern-.1667em\lower.7ex\hbox{E}\kern-.125emX}}

\graphicspath{{figures/}}

\begin{document}

\title{A Comparative Study of Encoding Strategies for Quantum Convolutional Neural Networks}

\author{Xingyun~Feng%
\thanks{Xingyun Feng is with Northwest University, Xi'an (e-mail: fengxingyun@stumail.nwu.edu.cn).}%
}

\markboth{X. Feng: Encoding Strategies for QCNNs}%
{X. Feng: Encoding Strategies for QCNNs}

\maketitle

\begin{abstract}
Quantum convolutional neural networks (QCNNs) offer a promising architecture for near-term quantum machine learning by combining hierarchical feature extraction with modest parameter growth. However, any QCNN operating on classical data must rely on an encoding scheme to embed inputs into quantum states, and this choice can dominate both performance and resource requirements. This work presents an implementation-level comparison of three representative encodings---Angle, Amplitude, and a Hybrid phase/angle scheme---for QCNNs under depolarizing noise. We develop a fully differentiable PyTorch--Qiskit pipeline with a custom autograd bridge, batched parameter-shift gradients, and shot scheduling, and use it to train QCNNs on downsampled binary variants of MNIST and Fashion-MNIST at $4\times 4$ and $8\times 8$ resolutions.

Our experiments reveal regime-dependent trade-offs. On aggressively downsampled $4\times 4$ inputs, Angle encoding attains higher accuracy and remains comparatively robust as noise increases, while the Hybrid encoder trails and exhibits non-monotonic trends. At $8\times 8$, the Hybrid scheme can overtake Angle under moderate noise, suggesting that mixed phase/angle encoders benefit from additional feature bandwidth. Amplitude-encoded QCNNs are sparsely represented in the downsampled grids but achieve strong performance in lightweight and full-resolution configurations, where training dynamics closely resemble classical convergence. Taken together, these results provide practical guidance for choosing QCNN encoders under joint constraints of resolution, noise strength, and simulation budget.
\end{abstract}


\section{Introduction}

Quantum machine learning (QML) seeks computational advantages by embedding learning tasks in quantum circuits whose expressivity derives from superposition and entanglement. Among QML models, Quantum Convolutional Neural Networks (QCNNs) are particularly attractive for near-term experiments: their hierarchical structure enables multiscale feature extraction with comparatively modest parameter growth, and their pooling pattern mirrors classical CNNs while reducing the active register across layers. Yet, before any variational circuit can learn, classical data must be embedded into quantum states. The choice of encoding is therefore pivotal---it governs not only accuracy but also resource footprint, trainability, and robustness under noise. Recent survey work highlights that encoding choices can become a dominant bottleneck for scalability and robustness in QML pipelines \cite{b1}, while theoretical analyses explicitly link feature maps to the expressive power of variational circuits \cite{b2}.

This work studies three representative encodings for QCNNs: Angle, Amplitude, and a Hybrid scheme that mixes phase and angle injections. Angle encoding maps features to single-qubit rotations and is shallow and hardware-amenable; Amplitude encoding prepares a normalized vector directly in the state amplitudes, achieving qubit efficiency at the cost of deeper state preparation; Hybrid encoding seeks a balance by combining phase and angle channels with light entanglement at the encoder. While these options are well known, systematic, side-by-side evidence in a unified implementation---especially under explicit noise models and realistic simulation budgets---remains limited. Closest to our study are comparative works on classical-to-quantum mappings \cite{b3}, which evaluate basis, angle, and amplitude encodings in hybrid pipelines. In contrast, we focus on QCNNs, explicit depolarizing-noise models, and extreme downsampling regimes, providing an architecture-specific, noise-aware comparison.

We present a reproducible simulation study of encoding strategies for QCNNs on downsampled binary variants of MNIST and Fashion-MNIST. The evaluation reflects practical constraints: (i) inputs are reduced to $4\times 4$ pixels for the main study and $8\times 8$ for replication; (ii) binary class pairs are chosen by a data-driven separability heuristic (pairwise L2 distance between class means) to ensure meaningful discrimination at extreme compression; (iii) depolarizing noise is applied at one- and two-qubit layers, and finite-shot estimation is used with a shot-scheduling policy to accelerate training. All results are derived from logged evaluations and configurations recorded in \texttt{experiment\_logs/}.

From an implementation standpoint, we develop an end-to-end differentiable pipeline that couples PyTorch optimization with Qiskit simulation. A custom autograd bridge implements a batched parameter-shift rule, consolidating all $\pm \pi/2$ evaluations for all parameters and all batch circuits into a single Estimator call in the backward pass. Together with shot scheduling for noisy simulations, this enables practical training and evaluation across a large grid of encoder/noise settings.

\section{Literature Review}

This section reviews the foundational concepts and prior work that underpin our study, focusing on the QCNN architecture and the data encoding techniques central to its application.

\subsection{Quantum Convolutional Neural Network (QCNN)}

The QCNN architecture was first introduced by Cong, Choi, and Lukin \cite{b4} as a quantum analogue to classical CNNs, designed for identifying features in quantum data. The model's structure is inspired by tensor networks, specifically the Multiscale Entanglement Renormalization Ansatz (MERA), which provides a hierarchical framework for feature extraction.

A QCNN consists of alternating layers of convolutional and pooling operations.
\begin{itemize}
  \item \textbf{Convolutional layers} apply parameterized two-qubit unitary gates to neighboring qubits. This operation creates entanglement and extracts local features. Critically, the variational parameters of these unitaries are shared across the layer, analogous to the shared weights of a classical convolutional filter.
  \item \textbf{Pooling layers} reduce the system's dimensionality. This is typically achieved by applying a controlled operation between two qubits and then discarding one of them (e.g., by measurement and reset), effectively pooling information from multiple qubits into a smaller subset.
\end{itemize}

This alternating structure systematically reduces the number of qubits while creating increasingly global correlations, enabling the QCNN to analyze data at multiple scales. Its logarithmic parameter count makes it a highly efficient variational quantum algorithm, well-suited for the constraints of NISQ-era devices. Beyond the original QCNN proposal \cite{b4} and quanvolutional hybrids that embed quantum patches into classical CNNs \cite{b5}, more recent hybrid quantum--classical--quantum convolutional architectures \cite{b6} interleave quantum filters, shallow classical convolutions, and trainable variational quantum classifiers, further blurring the boundary between encoders and ans\"atze.

\begin{figure*}[t!]
\centering
\vspace*{5pt}
\includegraphics[width=\textwidth,height=0.6\textheight,keepaspectratio]{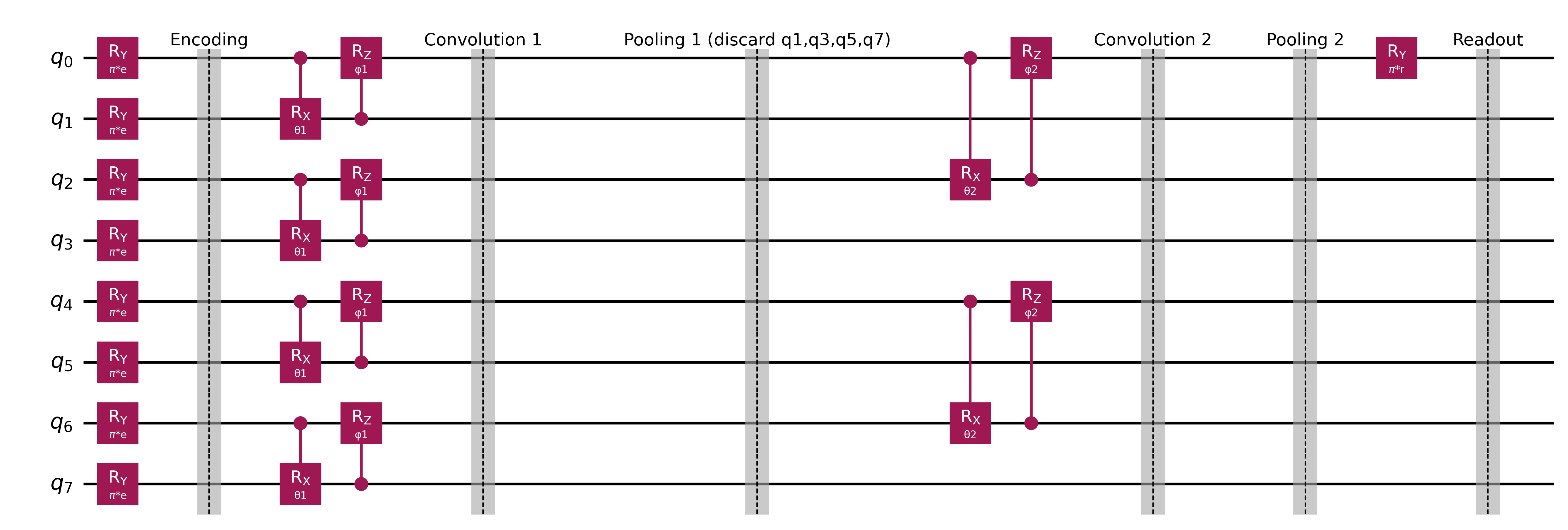}
\caption{Schematic overview of the QCNN architecture used in this work.}
\label{fig:qcnn_overview}
\end{figure*}

\subsection{Data Encoding Methods}

The performance of a QML model on classical data is heavily dependent on the chosen feature map \cite{b1}. This choice represents a critical trade-off between qubit resources, circuit depth, and the amount of information encoded. Our study focuses on three prominent methods.

\subsubsection{Angle encoding}
Also known as rotation encoding, this is one of the most direct methods for embedding classical data. As utilized in quanvolutional networks by Henderson \emph{et al.} \cite{b5}, each classical feature $x_j$ (e.g., a normalized pixel value) is mapped to the rotation angle of a single-qubit gate. For a feature vector $x$, the encoding can be expressed as
\begin{equation}
|\psi(x)\rangle = \bigotimes_{j=1}^{n} R_Y(\pi x_j)|0\rangle.
\label{eq:angle_encoding}
\end{equation}
For a patch of size $k\times k$ pixels, this requires $n=k^2$ qubits. This local approach is robust and simple to implement, requiring only shallow circuit depth per feature, making it highly compatible with near-term hardware.

\subsubsection{Amplitude encoding}
This technique offers exponential compression by embedding a $d$-dimensional normalized classical vector $x$ into the amplitudes of a quantum state using just $n=\lceil\log_2(d)\rceil$ qubits. The resulting state is
\begin{equation}
|x\rangle = \frac{1}{\|x\|} \sum_{i=0}^{d-1} x_i |i\rangle,
\label{eq:amplitude_encoding}
\end{equation}
where $\|x\|$ is the L2 norm of the vector. While exceptionally efficient in qubit count, its primary drawback is the state-preparation cost. Generating an arbitrary state with this method can require a circuit depth that scales polynomially with the vector size $d$, posing a significant challenge for NISQ devices susceptible to decoherence and gate errors.

\subsubsection{Hybrid encoding}
To mediate the trade-offs between Angle and Amplitude encoding, sophisticated schemes that combine multiple encoding primitives have been proposed. These methods often integrate the encoding and processing steps. For instance, prior work has explored strategies that combine Angle and Phase encoding ($R_Y$ and $R_Z$ gates) to store more information on each qubit, as well as schemes where data parameterizes multi-qubit entangling gates \cite{b8}. Our work implements a similar integrated strategy where features from an image patch are used to parameterize a sequence of single-qubit rotations and two-qubit entangling gates, thereby embedding the data while simultaneously creating entanglement. Conceptually, this design is related to data re-uploading architectures \cite{b9}, which alternate data-dependent rotations with trainable unitaries to boost expressivity. This offers a balanced compromise between qubit efficiency and circuit depth.

\begin{figure*}[t!]
\centering
\vspace*{5pt}
\includegraphics[width=\textwidth,height=0.6\textheight,keepaspectratio]{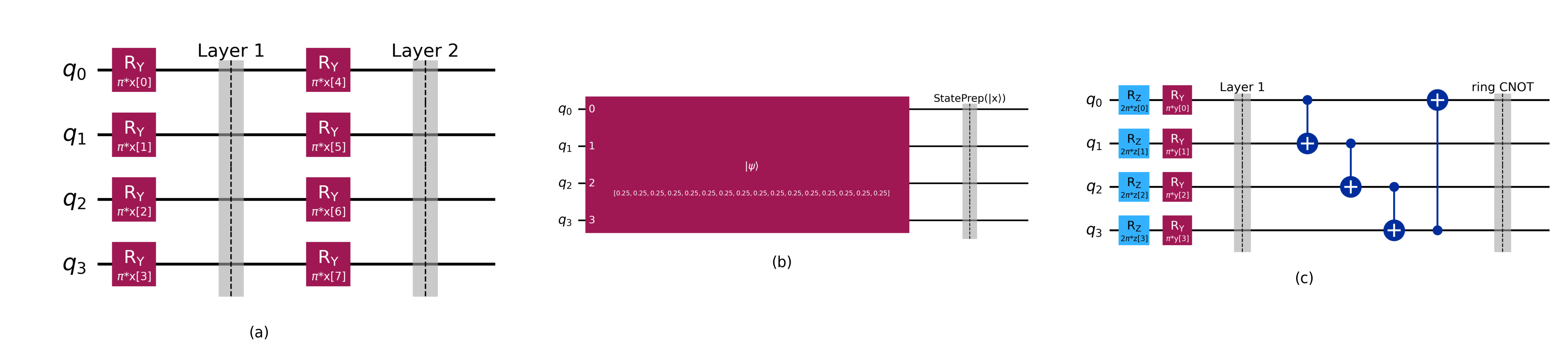}
\caption{Circuit-level schematic of the Angle, Amplitude, and Hybrid encoding schemes implemented in this study.}
\label{fig:encoding_schematic}
\end{figure*}

Despite the individual merits of these techniques, a practical, side-by-side comparison of their impact on a QCNN's classification performance, training dynamics, and resource costs within a unified experimental setup remains an open area of investigation. Our work complements theoretical analyses that link feature maps to the expressive power of variational models \cite{b2} and empirical comparisons of classical-to-quantum mappings in hybrid pipelines \cite{b3} by providing a concrete, QCNN-based empirical comparison under depolarizing noise.

\section{Methodology}

This section details the methodology for our comparative study of QCNNs. Our framework integrates PyTorch for classical optimization and Qiskit for quantum simulation, employing a custom differentiable quantum module to enable efficient, end-to-end training. The implementation has matured into an optimized, batched parameter-shift engine and a practical training workflow that supports noisy simulations, shot scheduling, and GPU acceleration on Qiskit Aer.

\subsection{System Architecture}

Our hybrid pipeline leverages PyTorch for dataset management and gradient-based optimization, while using Qiskit for quantum circuit construction and execution. At the core of this integration is a Qiskit Aer \texttt{Estimator} primitive, which can be configured for ideal or noisy simulations on either CPU or GPU backends.

A single training step processes a mini-batch of classical data through a quantum--classical computational graph. First, classical data samples are encoded into quantum states. These states are then processed by a parameterized QCNN ansatz. The expectation value of an observable---specifically, the Pauli-$Z$ operator on the final remaining qubit---is measured. This scalar expectation value is fed into a classical linear layer to produce class logits. Finally, gradients are computed analytically via the parameter-shift rule and are propagated back to update both the quantum and classical parameters.

\subsection{QCNN Model and Data Encoding}

\subsubsection{QCNN Architecture}

Our canonical QCNN model follows the hierarchical design proposed by Cong \emph{et al.} \cite{b4}. It consists of a stack of convolutional layers interleaved with pooling layers that progressively halve the active register size until a single qubit remains. Each convolutional layer applies a translationally invariant two-qubit unitary, implemented with single-qubit rotations followed by a CX entangling gate. The pooling layer also uses a CX gate to transfer information to the surviving qubits before the others are discarded. The final readout is the $Z$-expectation of the last qubit.

As a variant for ablation studies, we also implement a \texttt{QCNNDeep} model, which features a richer hypothesis class by incorporating additional RY, RZ, and RZZ gates into each convolutional block at the cost of more parameters and deeper circuits.

\subsubsection{Data Encoding Schemes}

Our study compares three distinct data encoding strategies. \textbf{Angle encoding} offers a direct and hardware-friendly approach by mapping each classical feature to the rotation of a single-qubit gate, requiring a circuit depth that is shallow and independent of the input data. In contrast, \textbf{Amplitude encoding} provides exponential qubit efficiency by embedding an entire $2^n$-dimensional feature vector into the amplitudes of an $n$-qubit state using an \texttt{Initialize} instruction, though at the cost of a more complex state-preparation circuit. To balance these trade-offs, we also evaluate a \textbf{Hybrid encoding} scheme, which uses feature values to parameterize a sequence of both single-qubit (RZ, RY) and two-qubit (CX) gates, thereby creating entanglement within the encoding layer itself.

\subsection{Datasets and Preprocessing}

We use the standard MNIST and Fashion-MNIST datasets. The preprocessing pipeline is tailored to the encoding scheme. For \textbf{Amplitude encoding}, the full $28\times 28$ image is flattened, zero-padded to match the dimension of the qubit register (e.g., 1024 for 10 qubits), and L2-normalized. These processed tensors are saved offline to \texttt{data/processed/} for efficient loading. For \textbf{Angle and Hybrid encoding}, a patch of the image (e.g., $8\times 8$ for 64 features) is extracted on-the-fly during data loading. For large experimental sweeps, we use pre-downsampled datasets (e.g., $4\times 4$ or $8\times 8$) to accelerate I/O and ensure consistency. For binary classification tasks, a subset of labels is selected and remapped to $\{0,1\}$ for compatibility with the cross-entropy loss function.

\subsection{Differentiable Training Framework}

A central challenge in this work was to create a fully differentiable framework that supports all encoding schemes while remaining highly efficient. High-level library abstractions like \texttt{TorchConnector} are incompatible with the \texttt{Initialize} instruction required for amplitude encoding, as the latter requires concrete data at circuit-construction time. To resolve this, we engineered a custom \texttt{torch.autograd.Function}. This module, which underpins our \texttt{QCNNOptimized} model, unifies all three encodings and implements a batched parameter-shift rule for efficient gradient computation.

The entire differentiable workflow, from data encoding to gradient computation, is illustrated below.

\begin{figure}[t!]
\centering
\vspace*{5pt}
\includegraphics[width=0.95\columnwidth,height=0.6\textheight,keepaspectratio]{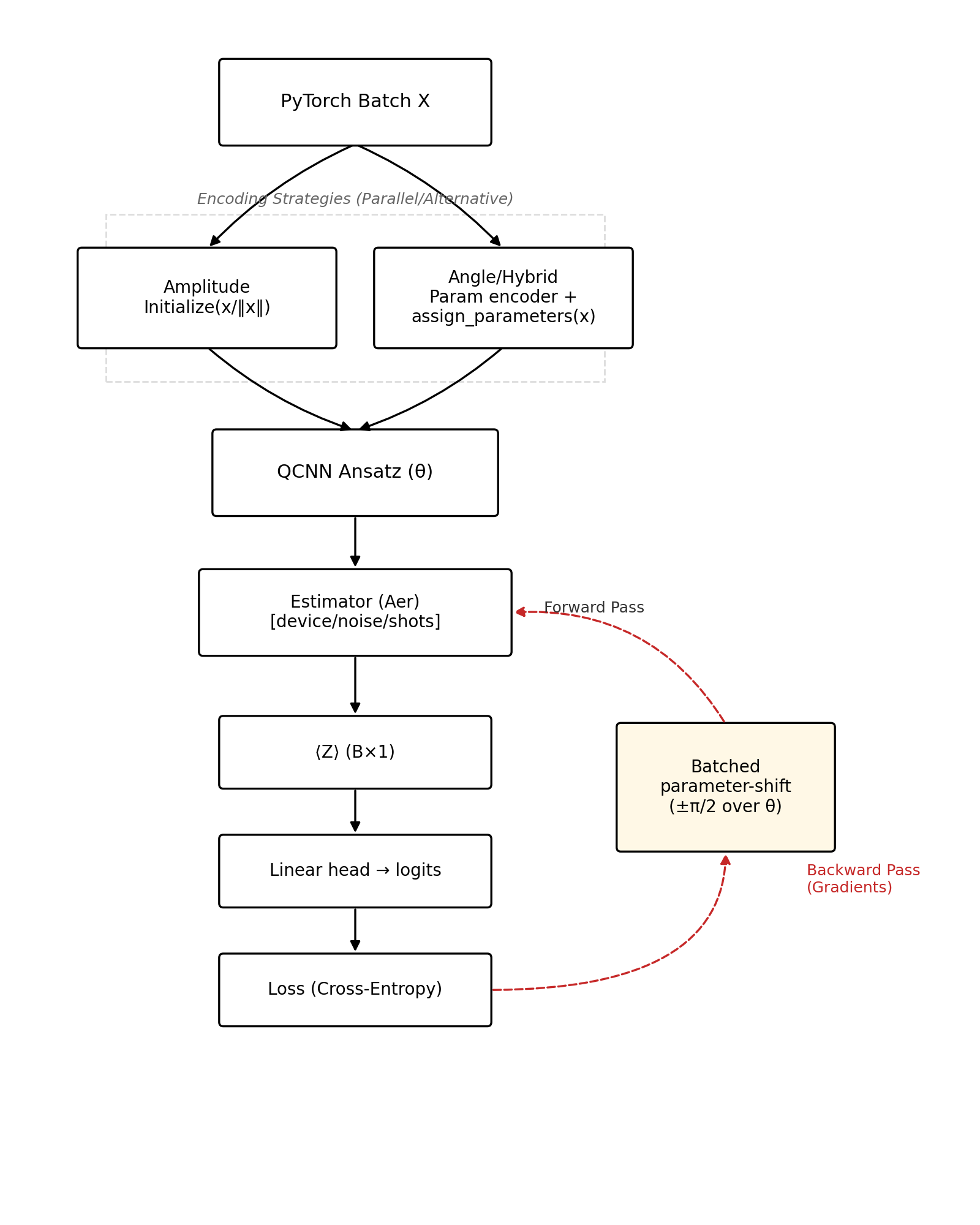}
\caption{End-to-end parameter and data flow from classical inputs through encoding, QCNN processing, and gradient back-propagation.}
\label{fig:param_flow}
\end{figure}

\subsubsection{Backward pass via batched parameter-shift}

To compute gradients for the $P$ trainable ansatz weights $\theta$, the parameter-shift rule is applied. The gradient for each parameter $\theta_i$ is given by
\begin{equation}
\frac{\partial \langle E \rangle}{\partial \theta_i} =
\frac{1}{2}\left(\langle E(\theta_i + \pi/2) \rangle - \langle E(\theta_i - \pi/2) \rangle \right).
\label{eq:param_shift}
\end{equation}
A naive implementation would require $2P$ separate \texttt{Estimator} calls per batch. We substantially reduce this overhead by batching all gradient computations into a single \texttt{Estimator} call.

\begin{figure*}[t!]
\centering
\vspace*{5pt}
\includegraphics[width=\textwidth,height=0.6\textheight,keepaspectratio]{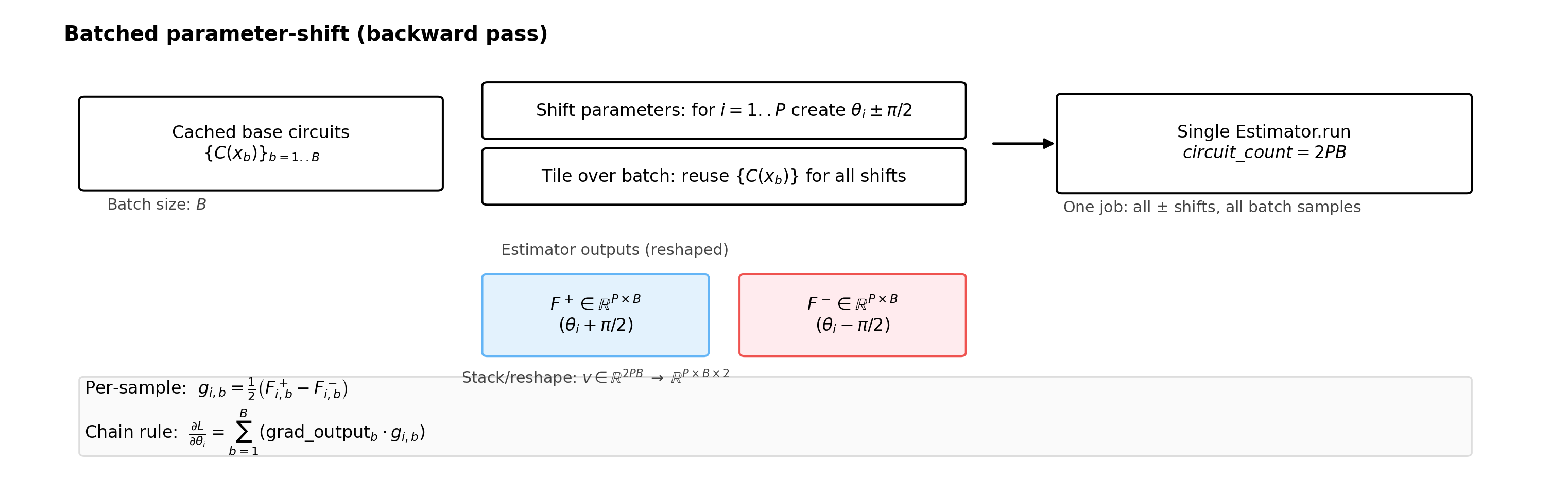}
\caption{Batched parameter-shift evaluation, aggregating all $\pm \pi/2$ weight shifts and all batch circuits into a single Estimator call.}
\label{fig:batched_shift}
\end{figure*}

\subsection{Simulation and Training Strategy}

\subsubsection{Noise simulation}

All experiments are conducted using the Qiskit \texttt{AerEstimator}. To assess model robustness, we configure it with a \texttt{NoiseModel} that applies a depolarizing channel to quantum gates. The model uses distinct error probabilities for single-qubit gates ($p_1$) and two-qubit gates ($p_2$), reflecting the higher error rates of entangling operations on real hardware.

\subsubsection{Training acceleration with shot scheduling}

In noisy simulations, the statistical variance of expectation values is inversely related to the number of shots. A fixed, high shot count is computationally wasteful in early training epochs where only a coarse gradient direction is needed. Conversely, a low shot count can hinder convergence in later epochs where high precision is required.

To balance this trade-off, we implement a shot scheduling strategy. At the start of each epoch $t\in[0,T-1]$, the number of shots is dynamically computed based on a predefined schedule, and the \texttt{AerEstimator} is rebuilt with this new value. We support several schedules, including a linear ramp from \texttt{min\_shots} to \texttt{max\_shots}:
\begin{equation}
\begin{aligned}
\mathrm{shots}_t &=
\operatorname{round}\Bigl(\mathrm{min\_shots} \\
&\quad + \frac{t}{T-1}\bigl(\mathrm{max\_shots}-\mathrm{min\_shots}\bigr)\Bigr).
\end{aligned}
\label{eq:shot_schedule}
\end{equation}
This approach reduces total wall-clock training time without compromising final model accuracy. For consistent reporting, we optionally run a brief evaluation after each epoch using a fixed, high shot count to obtain stable performance metrics.

\subsubsection{Experimental procedure}

All models are trained end-to-end using the Adam optimizer to minimize a cross-entropy loss function. The \texttt{QCNNOptimized} model serves as our primary architecture. During training, we save the last and best-performing model checkpoints, recording the configuration, epoch, accuracy, loss, and (for noisy runs) the number of shots used for that epoch to ensure reproducibility.

\subsubsection{Limitations and threats to validity}

Our methodology is subject to several limitations. First, the depolarizing noise model, while standard, is a simplification and does not capture all error sources of a specific quantum device. Second, the parameter-shift rule, while exact, scales linearly with the number of parameters, which may become a bottleneck for much larger models. Finally, our use of an ideal \texttt{Initialize} instruction for amplitude encoding abstracts away the significant circuit depth that would be required to implement it on real hardware. Hybrid classifiers with data re-uploading have already been demonstrated on small superconducting quantum devices \cite{b10}, suggesting that structurally similar encoder designs could be ported to hardware once more realistic noise and calibration models are taken into account.

\section{Experiments and Results}

This section presents the empirical comparison of Angle, Amplitude, and Hybrid encoding schemes for QCNNs under depolarizing noise. The experiments focus on binary classification tasks using downsampled versions of the MNIST and Fashion-MNIST datasets, designed to probe the performance and robustness of each encoding within a resource-constrained simulation environment. All numerical results are read directly from the logged evaluations and configuration records in \texttt{experiment\_logs/}.

\subsection{Experimental Setup and Evaluation Metrics}

We consider MNIST and Fashion-MNIST in binary form. To manage computational cost while retaining salient features, we use downsampled inputs at two resolutions: $4\times 4$ (16 features) as the main experimental regime, and $8\times 8$ (64 features) as a replication to assess the impact of more aggressive downsampling. For each dataset, the specific pair of classes for the binary task is chosen by computing pairwise L2 separability between class means and selecting a highly distinguishable pair (e.g., ``6'' vs ``7'' for MNIST), as implemented in \texttt{scripts/analyse\_downsampled\_l2.py}. The exact class pairs underlying each configuration are documented in Appendix~A.

All quantum models are instances of the \texttt{QCNNOptimized} architecture described in the Methodology section, using a batched parameter-shift rule for efficient gradient computation. Simulations are performed with the Qiskit Aer \texttt{Estimator}, configured with a depolarizing noise model characterized by error probabilities $p_1$ for single-qubit gates and $p_2$ for two-qubit gates. To accelerate noisy training, we employ the shot-scheduling strategy described earlier, dynamically adjusting the number of measurement shots across epochs.

The primary endpoint is test accuracy (\%). For each configuration we report the best evaluated test accuracy across checkpoints. When both noiseless (``none'') and high-noise conditions are present, we additionally consider the accuracy drop $\Delta_{\text{acc}}=\text{acc}_{\text{none}}-\text{acc}_{\text{high}}$. Resource footprint (qubit count, two-qubit gate count, and circuit depth) is summarized in the appendix; here we focus on accuracy and robustness. Classical CNN baselines trained on the same downsampled data are presented separately in Appendix~B, as their class pairs may differ from the QCNN tasks.

\subsection{Results on MNIST}

\subsubsection{$4\times 4$ resolution}

We first consider MNIST at a $4\times 4$ resolution (16 features), with the binary task ``6'' vs ``7''. The curated downsampled QCNN series provide complete coverage for Angle and Hybrid encodings across the considered noise levels; Amplitude-encoded QCNNs are not available in this regime and are therefore omitted from the main table.

\begin{table}[t]
\caption{MNIST $4\times 4$ accuracy (\%) versus noise level.}
\label{tab:mnist_4x4}
\centering
\setlength{\tabcolsep}{6pt}
\begin{tabular}{lcc}
\toprule
Noise & Angle & Hybrid \\
\midrule
none & -- & 46.875 \\
low  & 81.250 & 59.375 \\
mid  & 68.750 & 51.563 \\
high & 68.750 & 57.813 \\
\bottomrule
\end{tabular}
\end{table}

\noindent\emph{Note:} the noiseless (``none'') result for Angle is absent in the curated series. The Hybrid ``none'' entry is supplemented from a later series and is marked here for completeness.

Angle encoding achieves over 81\% accuracy in the low-noise setting and maintains relatively high performance as noise strength increases. Hybrid trails in absolute accuracy and exhibits a non-monotonic response to noise, reflecting the interaction between mixed phase/amplitude injection and the shallow QCNN at this feature budget.

\subsubsection{$8\times 8$ resolution}

To assess whether the $4\times 4$ downsampling excessively reduces image contrast, we replicate the study at $8\times 8$ (64 features). In this regime the logs provide values for Angle and Hybrid under low, mid, and high noise, predominantly with the class pair ``0'' vs ``1''.

\begin{table}[t]
\caption{MNIST $8\times 8$ accuracy (\%) versus noise level.}
\label{tab:mnist_8x8}
\centering
\setlength{\tabcolsep}{6pt}
\begin{tabular}{lcc}
\toprule
Noise & Angle & Hybrid \\
\midrule
low  & 75.000 & 56.250 \\
mid  & 56.250 & 75.000 \\
high & 68.750 & 62.500 \\
\bottomrule
\end{tabular}
\end{table}

\noindent\emph{Note:} noiseless (``none'') results are not available for this resolution in the curated downsampled QCNN series.

At this higher resolution, a notable crossover occurs: under mid-level noise the Hybrid encoder attains 75.00\% accuracy versus 56.25\% for Angle, suggesting that the Hybrid scheme can better exploit the additional feature bandwidth. Angle remains competitive at low and high noise.

\subsection{Results on Fashion-MNIST}

To probe generality beyond MNIST, we apply the same protocol to Fashion-MNIST.

\subsubsection{$4\times 4$ resolution}

At $4\times 4$ resolution, the curated downsampled QCNN logs provide a clear performance trend for the Hybrid encoder on the ``T-shirt/top'' vs ``Trouser'' (4 vs 5) task; Angle and Amplitude encodings are not consistently available in this regime.

\begin{table}[t]
\caption{Fashion-MNIST $4\times 4$ accuracy (\%) versus noise level.}
\label{tab:fashion_4x4}
\centering
\setlength{\tabcolsep}{6pt}
\begin{tabular}{lc}
\toprule
Noise & Hybrid \\
\midrule
low  & 93.750 \\
mid  & 75.000 \\
high & 75.000 \\
\bottomrule
\end{tabular}
\end{table}

Hybrid achieves excellent accuracy (93.75\%) in the low-noise setting. Performance decreases under mid and high noise but remains substantially above chance, indicating that the QCNN can still extract discriminative structure from a highly compact 16-pixel representation.

\subsubsection{$8\times 8$ resolution}

At $8\times 8$, the curated Hybrid entries cover low and mid noise on the ``T-shirt/top'' vs ``Sandal'' (0 vs 7) task; ``high'' and ``none'' are absent in the downsampled QCNN logs.

\begin{table}[t]
\caption{Fashion-MNIST $8\times 8$ accuracy (\%) versus noise level.}
\label{tab:fashion_8x8}
\centering
\setlength{\tabcolsep}{6pt}
\begin{tabular}{lc}
\toprule
Noise & Hybrid \\
\midrule
low  & 75.000 \\
mid  & 62.500 \\
\bottomrule
\end{tabular}
\end{table}

The drop from low to mid noise mirrors the qualitative trend observed on MNIST, although absolute figures differ due to dataset characteristics and class-pair choices.

\subsubsection{Aggregate view}

\begin{figure}[t!]
\centering
\vspace*{5pt}
\includegraphics[width=0.95\columnwidth,height=0.6\textheight,keepaspectratio]{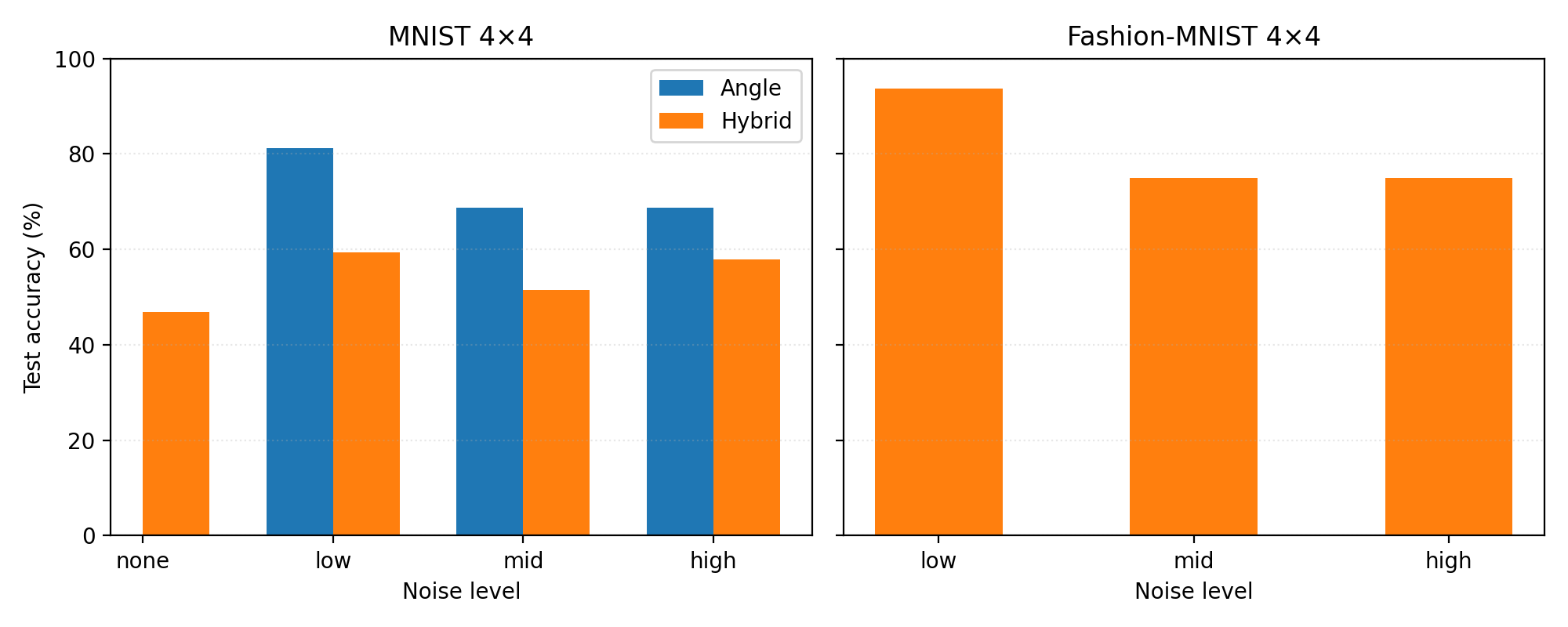}
\caption{Test accuracy versus depolarizing noise level for QCNN encoders on $4\times 4$ downsampled MNIST (Angle and Hybrid) and Fashion-MNIST (Hybrid).}
\label{fig:bar_4x4}
\end{figure}

\begin{figure}[t!]
\centering
\vspace*{5pt}
\includegraphics[width=0.95\columnwidth,height=0.6\textheight,keepaspectratio]{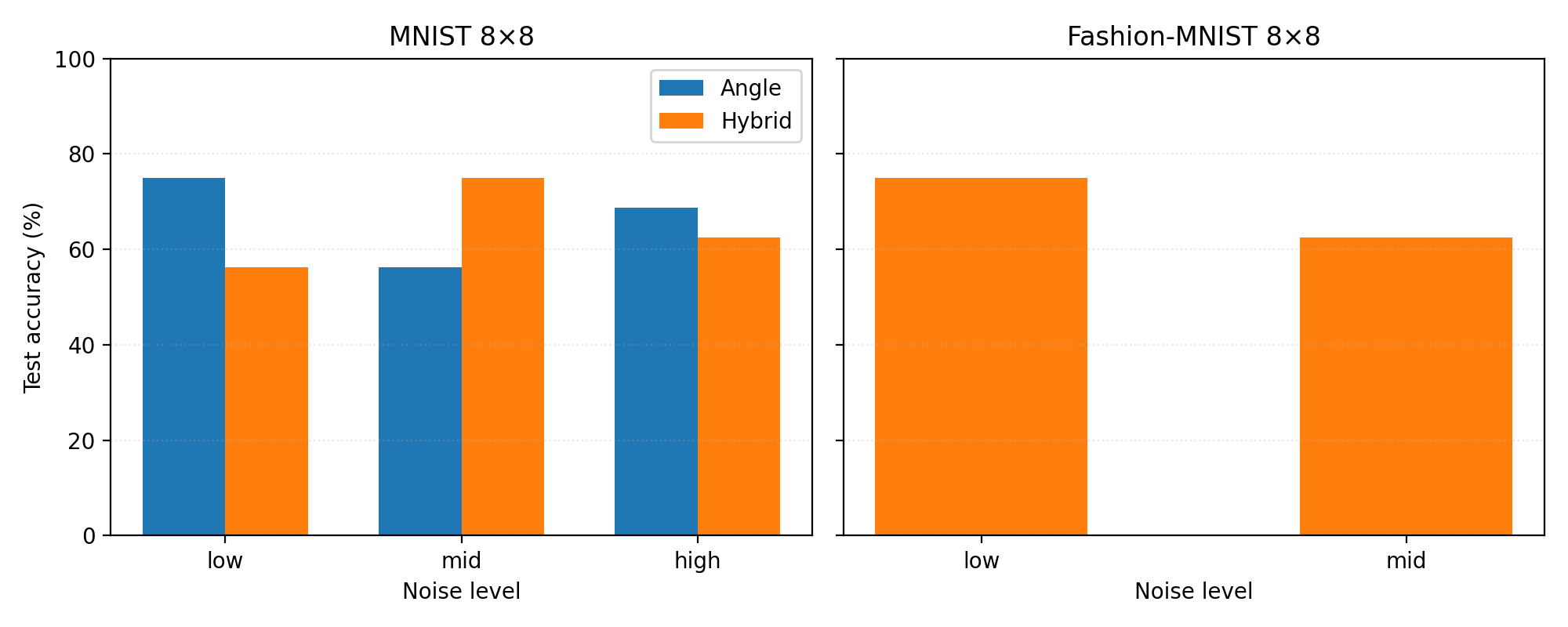}
\caption{Test accuracy versus depolarizing noise level for QCNN encoders on $8\times 8$ downsampled MNIST (Angle and Hybrid) and Fashion-MNIST (Hybrid).}
\label{fig:bar_8x8}
\end{figure}

Figure~\ref{fig:bar_4x4} summarizes the $4\times 4$ results on MNIST and Fashion-MNIST, while Fig.~\ref{fig:bar_8x8} depicts the corresponding $8\times 8$ replications. The bar charts visually reinforce the numerical trends in Tables~\ref{tab:mnist_4x4}--\ref{tab:fashion_8x8}: Angle is competitive and relatively stable at extreme compression on MNIST, whereas Hybrid dominates at moderate noise when a larger feature budget is available; on Fashion-MNIST, Hybrid performs strongly in the low-noise regime and degrades under stronger noise while remaining well above chance.

\subsection{Training Dynamics and Additional Experiments}

In the $4\times 4$ downsampled setting with noisy finite-shot evaluation, per-epoch test accuracies on MNIST exhibit considerable variability across runs and checkpoints, making it difficult to extract a clear notion of convergence from single training curves. Rather than over-emphasizing these noisy trajectories, we treat the best-checkpoint accuracies in Tables~\ref{tab:mnist_4x4}--\ref{tab:fashion_8x8} as the primary indicators of performance in the extreme-compression regime.

\begin{figure}[t!]
\centering
\vspace*{5pt}
\includegraphics[width=0.95\columnwidth,height=0.6\textheight,keepaspectratio]{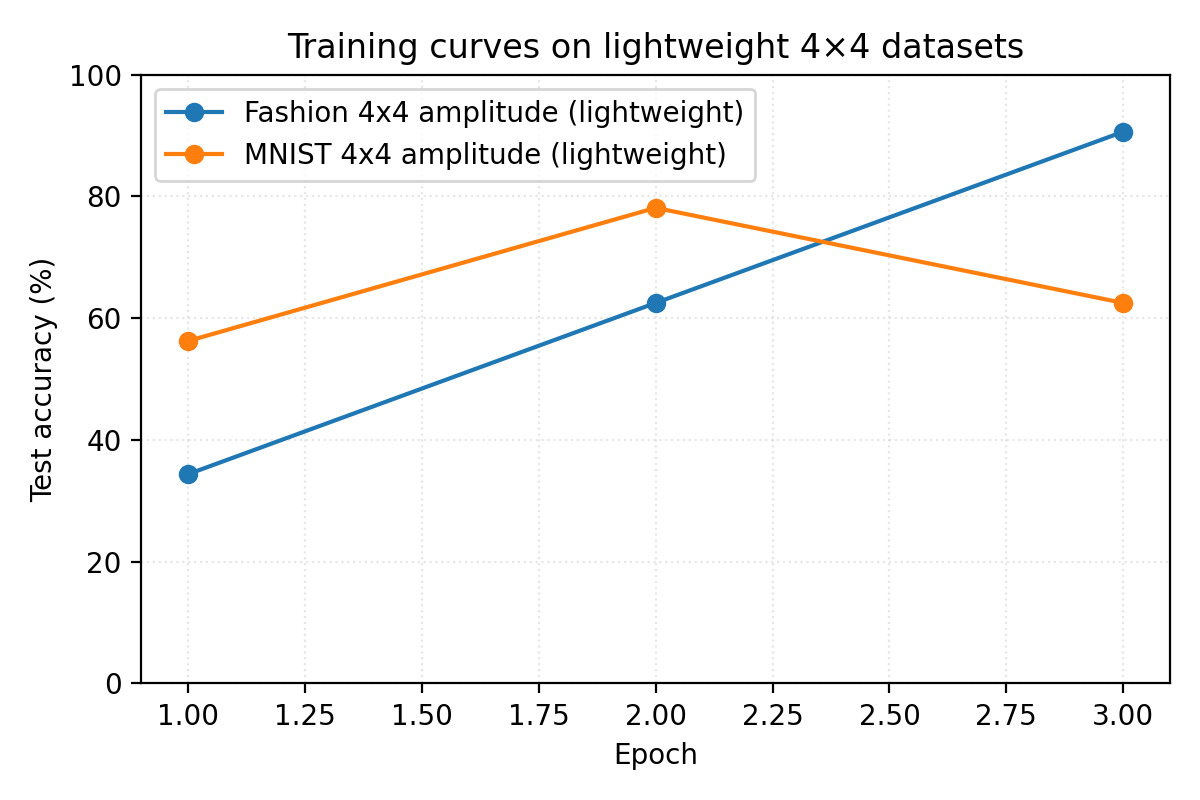}
\caption{Test accuracy versus epoch for lightweight QCNNs on MNIST and Fashion-MNIST ($4\times 4$ amplitude encoding).}
\label{fig:lightweight_curves}
\end{figure}

To demonstrate that the overall framework admits more conventional convergence behavior under less aggressive settings, we also examine two complementary scenarios. First, we consider lightweight configurations trained on larger effective datasets. Fig.~\ref{fig:lightweight_curves} shows training curves for a lightweight amplitude-encoded QCNN on Fashion-MNIST and on MNIST, both using $4\times 4$ patches but with significantly more training samples per epoch than the extreme downsampled experiments. For each configuration and epoch, we plot the maximum evaluated test accuracy over all checkpoints stored at that epoch in the corresponding evaluation logs.

On Fashion-MNIST, the amplitude-encoded lightweight QCNN exhibits a clear upward trend, with accuracy rising from 34.38\% at epoch 1 to 62.50\% at epoch 2 and 90.62\% at epoch 3. On MNIST, the corresponding configuration improves from 56.25\% at epoch 1 to 78.13\% at epoch 2 before settling near 62.50\% at epoch 3. These trajectories illustrate that the optimization pipeline can display familiar convergence patterns when provided with more informative inputs and larger effective sample sizes.

\begin{figure}[t!]
\centering
\vspace*{5pt}
\includegraphics[width=0.95\columnwidth,height=0.6\textheight,keepaspectratio]{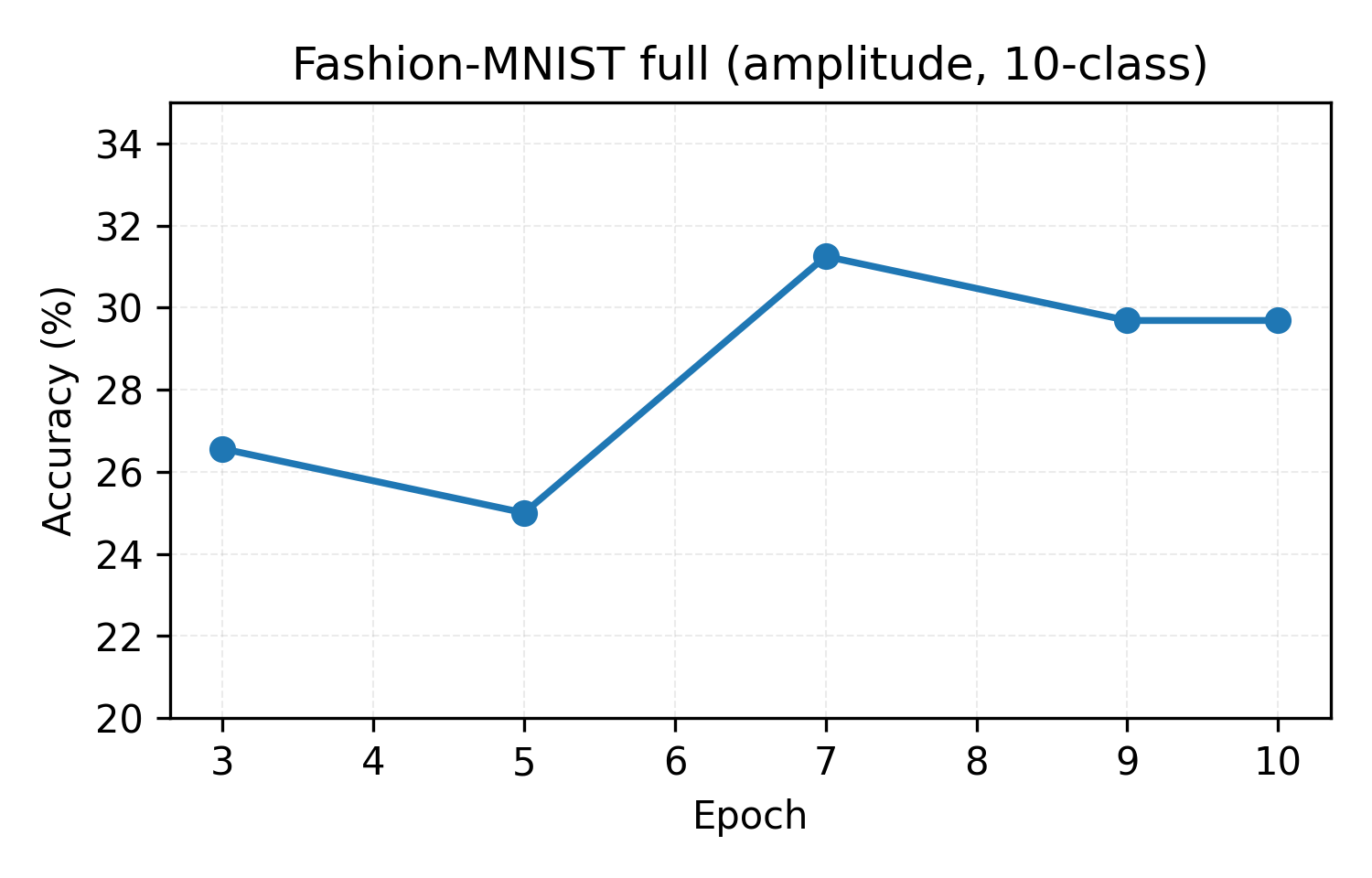}
\caption{Test accuracy versus epoch for a QCNN with amplitude encoding on full-resolution Fashion-MNIST (10-class classification).}
\label{fig:fashion_fullres_curve}
\end{figure}

Second, we examine a full-resolution configuration on Fashion-MNIST using Amplitude encoding and the complete ten-class task. Fig.~\ref{fig:fashion_fullres_curve} shows the training curve for this experiment; again, for each epoch we plot the maximum evaluated test accuracy across all checkpoints written at that epoch.

\begin{figure}[t!]
\centering
\vspace*{5pt}
\includegraphics[width=0.95\columnwidth,height=0.6\textheight,keepaspectratio]{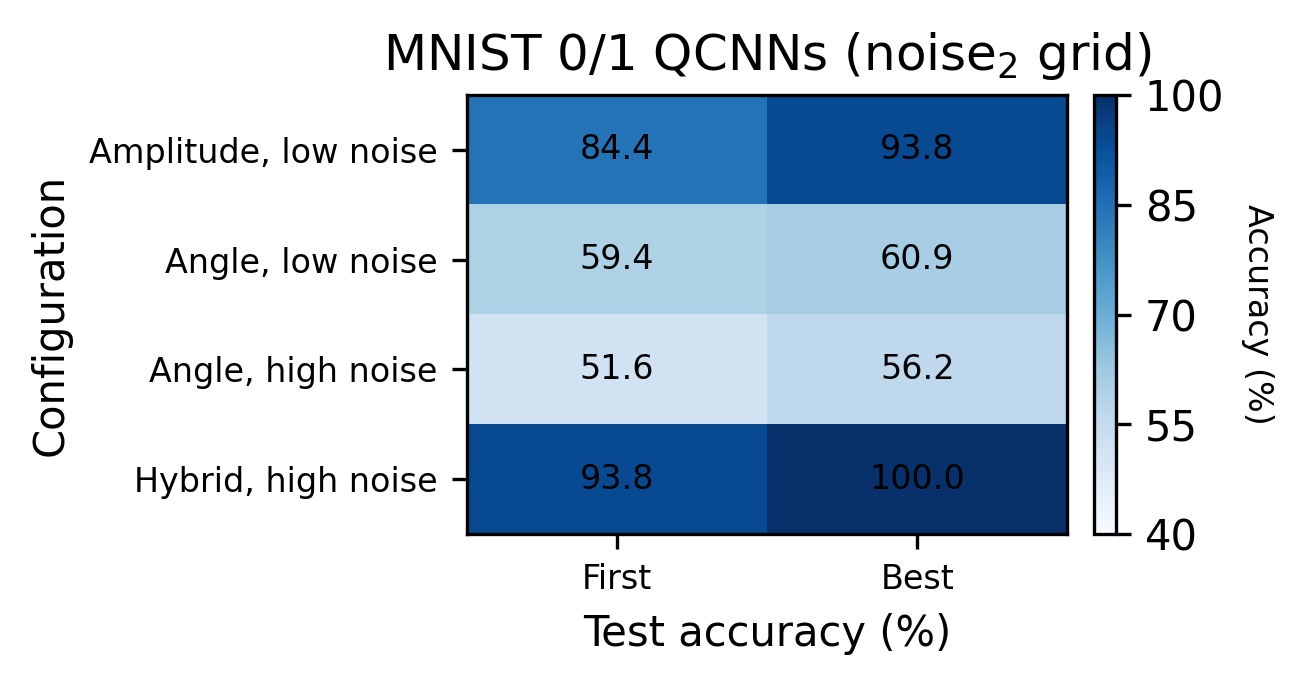}
\caption{First and best evaluated test accuracies for MNIST 0/1 QCNNs under selected \texttt{noise\_2} settings, visualized as a $4\times 2$ heatmap (Amplitude--low noise, Angle--low and high noise, Hybrid--high noise).}
\label{fig:mnist_heatmap}
\end{figure}

To illustrate how encoding choice affects attainable performance on a full-resolution binary task, Fig.~\ref{fig:mnist_heatmap} presents selected configurations from the \texttt{noise\_2} grid: the test accuracy at the first evaluated checkpoint and at the best checkpoint recorded in the logs. Amplitude under low noise and Hybrid under high noise achieve high final accuracies (above 90\% and up to 100\%), while Angle configurations under low and high noise remain in the 50--60\% range.

\subsection{Noise Sensitivity, Scaling, and Limitations}

Where both noiseless and high-noise conditions are present, we can examine none$\rightarrow$high deltas. On MNIST $4\times 4$ Hybrid, accuracy increases from 46.875\% (none; supplemented from a later series) to 57.813\% (high), yielding a $\Delta_{\text{acc}}$ of $-10.94$ percentage points. This counter-intuitive behavior is likely an artifact of stochasticity in the small-model, small-data regime combined with finite-shot estimation. Rather than over-interpreting single cells, we emphasize the broader patterns apparent in Tables~\ref{tab:mnist_4x4}--\ref{tab:fashion_8x8} and Figs.~\ref{fig:bar_4x4}--\ref{fig:mnist_heatmap}. In particular, the combination of downsampling, noise, and limited epochs can induce non-monotonic trends that should be revisited under larger-scale simulations or hardware experiments.

In attempting to stabilize training on $4\times 4$ inputs, we also experimented with simple preprocessing heuristics for Angle and Hybrid encoders: L2-normalizing each flattened patch and applying a global multiplicative scale to the encoder angles. Empirically, such normalization and scaling tended to improve best accuracies in low- and mid-noise settings for some configurations, while offering little benefit or even reducing accuracy under strong noise. A plausible interpretation is that, before normalization, coarse statistics such as overall brightness or energy of the patch can themselves serve as robust discriminative signals; when these are attenuated by normalization, the model is forced to rely more heavily on fine-grained structure, which is more susceptible to depolarizing noise. For Amplitude encoding, by contrast, the state-preparation step already enforces unit L2 norm on the feature vector, so additional global scaling is mathematically absorbed by the normalization and does not affect the encoded quantum state.

These observations are empirical and configuration-dependent: scaling and normalization were adjusted alongside other hyperparameters within each noise grid, and we did not design a dedicated single-variable ablation. We therefore view these heuristics as practical stabilization tools for the most aggressive downsampling regime, rather than as primary drivers of the comparative conclusions on encoding strategies.

All numbers in this section are sourced directly from the evaluation logs and configuration catalog; full per-series tables and configuration paths are listed in Appendix~A for reproducibility.

\section{Conclusion}

This work has presented a unified, implementation-level comparison of three encoding strategies---Angle, Amplitude, and Hybrid---for QCNNs. On aggressively downsampled $4\times 4$ inputs, Angle encoding emerges as a competitive and comparatively robust option on MNIST, maintaining high accuracy across noise levels, while Hybrid trails and can exhibit non-monotonic behavior. When the feature budget is increased to $8\times 8$, Hybrid can overtake Angle under moderate noise, particularly on MNIST, indicating that mixed phase/angle encoders can better exploit additional spatial information. On Fashion-MNIST $4\times 4$ and $8\times 8$, Hybrid performs strongly at low noise and remains well above chance even as depolarizing noise increases, underscoring that the optimal encoding depends on both dataset and resolution.

From an implementation perspective, the study demonstrates that a custom PyTorch--Qiskit autograd bridge, combined with batched parameter-shift and shot scheduling, can support a large grid of noisy QCNN experiments without resorting to approximate gradients. Lightweight and full-resolution experiments confirm that, outside of the extreme $4\times 4$ compression regime, the same framework exhibits conventional convergence behavior, and that high accuracies are achievable for binary tasks on full-resolution MNIST and Fashion-MNIST. At the same time, the limited coverage of Amplitude-encoded QCNNs in the downsampled grids and the modest improvements under aggressive noise highlight that encoding comparisons are inherently configuration-dependent and must be interpreted in the context of resource constraints.

\section{Outlook}

Several directions follow naturally from this work. First, expanding the experimental coverage for Amplitude encoders---especially on intermediate resolutions and less aggressive downsampling---would help clarify their role beyond the current full-resolution baselines. Second, extending QCNN depth and qubit count while monitoring gradient scaling and optimization stability would connect the present study to broader questions about variational trainability. Third, applying the same encoding and training framework to additional datasets or real quantum hardware, with more realistic noise and calibration models, would test whether the observed Angle--Hybrid trade-offs persist beyond simulation. Related hybrid quantum--classical CNNs with data re-uploading filters \cite{b11} and QCQ-CNN architectures \cite{b6} pursue complementary directions in which quantum feature maps are interleaved with classical convolutions; our study keeps the convolutional stack fully quantum and varies only the encoder. Finally, more systematic ablations of preprocessing heuristics (including normalization and scaling) could turn the empirical stabilization tricks used here into more general guidelines for extreme-compression regimes.

\appendices

\section{Per-Series Downsampled QCNN Results (Traceability)}

This appendix lists per-series, per-configuration results for the downsampled binary tasks reported in the Experiments section. Each row corresponds to an entry curated directly from \texttt{experiment\_logs}, including the originating series and YAML path. Full CSVs are provided to avoid transcription error:
\begin{itemize}
  \item \path{paper_assets/data/table_series_qcnn_mnist_4x4.csv}
  \item \path{paper_assets/data/table_series_qcnn_mnist_8x8.csv}
  \item \path{paper_assets/data/table_series_qcnn_fashionmnist_4x4.csv}
  \item \path{paper_assets/data/table_series_qcnn_fashionmnist_8x8.csv}
\end{itemize}
For convenience, we reproduce compact excerpts below.

\begin{table*}[t]
\caption{MNIST $4\times 4$ (series detail; labels=6,7).}
\label{tab:series_mnist_4x4}
\centering
\setlength{\tabcolsep}{5pt}
\scriptsize
\begin{tabular}{lllll}
\toprule
Encoding & Noise & Acc (\%) & Series & Config \\
\midrule
angle  & low  & 81.250 & noise\_2.3.1 & \url{configs/noise_2.3.1/mnist_angle_noise_low.yaml} \\
angle  & mid  & 68.750 & noise\_2.3.1 & \url{configs/noise_2.3.1/mnist_angle_noise_mid.yaml} \\
angle  & high & 68.750 & noise\_2.3.1 & \url{configs/noise_2.3.1/mnist_angle_noise_high.yaml} \\
hybrid & none & 46.875 & noise\_2.6   & \url{configs/noise_2.6/mnist_hybrid_noise_none.yaml} \\
hybrid & low  & 59.375 & noise\_2.6   & \url{configs/noise_2.6/mnist_hybrid_noise_low.yaml} \\
hybrid & mid  & 51.563 & noise\_2.6   & \url{configs/noise_2.6/mnist_hybrid_noise_mid.yaml} \\
hybrid & high & 57.813 & noise\_2.6   & \url{configs/noise_2.6/mnist_hybrid_noise_high.yaml} \\
\bottomrule
\end{tabular}
\end{table*}

\begin{table*}[t]
\caption{MNIST $8\times 8$ (series detail; labels predominantly 0,1).}
\label{tab:series_mnist_8x8}
\centering
\setlength{\tabcolsep}{5pt}
\scriptsize
\begin{tabular}{lllll}
\toprule
Encoding & Noise & Acc (\%) & Series & Config \\
\midrule
angle  & low  & 75.000 & noise\_2.5.1 & \url{configs/noise_2.5.1/mnist_angle_noise_low.yaml} \\
angle  & mid  & 56.250 & noise\_2.5.1 & \url{configs/noise_2.5.1/mnist_angle_noise_mid.yaml} \\
angle  & high & 68.750 & noise\_2.5.1 & \url{configs/noise_2.5.1/mnist_angle_noise_high.yaml} \\
hybrid & low  & 56.250 & noise\_2.5.1 & \url{configs/noise_2.5.1/mnist_hybrid_noise_low.yaml} \\
hybrid & mid  & 75.000 & noise\_2.5.1 & \url{configs/noise_2.5.1/mnist_hybrid_noise_mid.yaml} \\
hybrid & high & 62.500 & noise\_2.5.1 & \url{configs/noise_2.5.1/mnist_hybrid_noise_high.yaml} \\
\bottomrule
\end{tabular}
\end{table*}

\begin{table*}[t]
\caption{Fashion-MNIST $4\times 4$ (series detail; labels=4,5).}
\label{tab:series_fashion_4x4}
\centering
\setlength{\tabcolsep}{5pt}
\scriptsize
\begin{tabular}{lllll}
\toprule
Encoding & Noise & Acc (\%) & Series & Config \\
\midrule
hybrid & low  & 93.750 & noise\_2.3.1 & \url{configs/noise_2.3.1/fashion_hybrid_noise_low.yaml} \\
hybrid & mid  & 75.000 & noise\_2.3.1 & \url{configs/noise_2.3.1/fashion_hybrid_noise_mid.yaml} \\
hybrid & high & 75.000 & noise\_2.3.1 & \url{configs/noise_2.3.1/fashion_hybrid_noise_high.yaml} \\
\bottomrule
\end{tabular}
\end{table*}

\begin{table*}[t]
\caption{Fashion-MNIST $8\times 8$ (series detail; labels=0,7).}
\label{tab:series_fashion_8x8}
\centering
\setlength{\tabcolsep}{5pt}
\scriptsize
\begin{tabular}{lllll}
\toprule
Encoding & Noise & Acc (\%) & Series & Config \\
\midrule
hybrid & low  & 75.000 & noise\_2.5.1 & \url{configs/noise_2.5.1/fashion_hybrid_noise_low.yaml} \\
hybrid & mid  & 62.500 & noise\_2.5.1 & \url{configs/noise_2.5.1/fashion_hybrid_noise_mid.yaml} \\
\bottomrule
\end{tabular}
\end{table*}

\section{CNN Baselines}

Classical CNN baselines are maintained separately to avoid conflating problem definitions when label pairs differ. Evaluations are recorded under \path{experiment_logs/cnn_eval_results.json}; per-config-to-checkpoint mappings are listed in \path{experiment_logs/config_checkpoint_catalog.md} (section ``CNN Baselines''). The baseline tables can be regenerated via the existing evaluator scripts; we omit them here to keep the main narrative focused on QCNN encoders.

\section{Reproducibility Files}

The curated inputs used to populate the Experiments section are included as CSV files:
\begin{itemize}
  \item \url{paper_assets/data/curated_best_downsampled.csv} (all datasets/sizes)
  \item \url{paper_assets/data/table_qcnn_mnist_4x4.csv}
  \item \url{paper_assets/data/table_qcnn_mnist_8x8.csv}
  \item \url{paper_assets/data/table_qcnn_fashionmnist_4x4.csv}
  \item \url{paper_assets/data/table_qcnn_fashionmnist_8x8.csv}
\end{itemize}
These files are produced by \url{paper_assets/scripts/curate_downsampled_experiments.py} from \texttt{experiment\_logs} JSONs and the referenced YAML configurations.

\section{Reporting Protocol and Data Curation}

To ensure transparency and traceability, all experimental results reported in the Experiments section adhere to a strict protocol.

\textbf{Data curation:} Every reported number is curated directly from the raw JSON logs in the \texttt{experiment\_logs/} directory. The selection and aggregation process is automated by the \url{paper_assets/scripts/curate_downsampled_experiments.py} script, which generates the final data tables. No manual post-processing or data imputation is performed beyond the rules defined in the script.

\textbf{Reporting rules:} For each experimental condition (dataset, input size, encoding, noise level), the tables in the Experiments section report the accuracy from the single best-performing checkpoint, where ``best'' is defined by the highest test accuracy achieved during training. When an experimental series lacks a specific noise condition (e.g., a ``none'' or noiseless run), the corresponding cell in the results table is either left blank or, if a comparable result from a different but compatible series (e.g., \texttt{noise\_2.6.3}) is available, it is used and explicitly noted in the text.

\textbf{Traceability:} The exact experimental series, configuration file (\texttt{.yaml}), and class pair used for every data point are documented in the per-series tables within this Appendix, providing a complete and traceable record of the results.

\section*{Acknowledgment}

The author thanks the open-source communities of PyTorch and Qiskit for their software and documentation.

\end{document}